\documentclass[
aps
,pra
,superscriptaddress
,showkeys
,twocolumn
,longbibliography
]{revtex4-1}

\usepackage{amsmath, amsthm, amssymb}
\usepackage{graphicx}
\usepackage{xcolor}
\usepackage{xargs}
\usepackage[colorlinks,citecolor=blue,urlcolor=blue]{hyperref}
\usepackage[utf8]{inputenc}
\usepackage{ragged2e}

\newcommand{\vv}[1]{\mathbf{#1}}

\newcommand{\vvk}[0]{\vv{k}}
\newcommand{\vvr}[0]{\vv{r}}
\newcommand{\vvv}[0]{\vv{v}}
\newcommand{\vvE}[0]{\vv{E}}

\newcommand{\vvB}[0]{\vv{B}}

\newcommand{\energy}[0]{\mathcal{E}}

\newcommand{\vx}[1]{v_x^{#1}}

\newcommand{\PPvvk}[1]{\frac{ \partial #1 }{ \partial \vvk } }

\def\Put(#1,#2)#3{\leavevmode\makebox(0,0){\put(#1,#2){#3}}}

\newcommand{\eig}[0]{\varepsilon_{k,n}}

\newcommand{\revision}[1]{\textcolor{black}{{#1}}}
\definecolor{purple}{rgb}{0.62, 0., 0.77}

\begin{document}

\title{Resonant Nonlinear Hall Effect in Two-Dimensional Electron Systems }
\author{Botsz Huang}
\thanks{These two authors contributed equally to this work.}
\affiliation{Department of Physics, National Cheng Kung University, Taiwan}
\author{Ali G. Moghaddam}
\thanks{These two authors contributed equally to this work.}
\affiliation{Department of Physics, Institute for Advanced Studies in Basic Sciences (IASBS), Zanjan 45137-66731, Iran}
\affiliation{Institute for Theoretical Solid State Physics, IFW Dresden, Helmholtzstr. 20, 01069 Dresden, Germany}
\author{Jorge I. Facio}\email{j.facio@ifw-dresden.de}
\affiliation{Institute for Theoretical Solid State Physics, IFW Dresden, Helmholtzstr. 20, 01069 Dresden, Germany}
\author{Ching-Hao Chang}\email{cutygo@phys.ncku.edu.tw}\affiliation{Department of Physics, National Cheng Kung University, Taiwan} \affiliation{Center for Quantum Frontiers of Research and Technology (QFort), National Cheng Kung University, Tainan 70101, Taiwan}

\date{\today} 

\begin{abstract}
We study the Hall conductivity of a two-dimensional electron gas under an inhomogeneous magnetic field $B(x)$. 
First, we prove using the quantum kinetic theory that an odd magnetic field can lead to a purely nonlinear Hall response. Second, considering a real-space magnetic dipole consisting of a sign-changing magnetic field and based on numerical semiclassical dynamics, we unveil a parametric resonance involving the cyclotron ratio and a characteristic  width of $B(x)$, which can greatly enhance the Hall response. Different from previous mechanisms that rely on the bulk Berry curvature dipole, the effect largely stems from boundary states associated with the real-space magnetic dipole. Our findings pave a new way to engineer current rectification and higher harmonic generation in two-dimensional materials having or not crystal inversion symmetry.
\end{abstract}

\maketitle

\section{Introduction}

It has been recently predicted that materials having time-reversal symmetry ($\Theta$) can present a Hall effect in the absence of a magnetic field if the crystal symmetries are sufficiently low \cite{inti_prl_2015,PhysRevB.92.235447,deyo2009semiclassical}.
This effect requires the inversion symmetry to be broken and is nonlinear in the applied electric field.
In the absence of disorder, the so-called Berry curvature dipole -- a measure of the average Berry curvature of a displaced Fermi surface -- sets the scale of the Hall response \cite{PhysRevB.99.155404,PhysRevLett.123.246602}.
Subsequent experimental confirmation in various systems \cite{ma2019observation,kang2019nonlinear,PhysRevLett.123.036806,shvetsov2019nonlinear,huang2020giant,ho2021hall} has further spurred interest in various directions including the search of systems where to study this phenomena \cite{PhysRevB.97.035158,PhysRevB.97.041101,PhysRevB.98.121109,PhysRevLett.121.266601,PhysRevLett.121.246403,PhysRevLett.121.266601,PhysRevLett.121.266601,PhysRevLett.123.196403,PhysRevLett.124.067203,PhysRevLett.125.046402,PhysRevB.102.245116,polini2018,Juridic2020,wawrzik2020infinite,carmine2021review,du2021perspective,malla2021emerging}, the effects of disorder \cite{du2019disorder,PhysRevB.100.165422,Sodemann2019,du2020quantum,isobe2020high} and the identification of possible technological profits, such as its usage for rectification or higher harmonic generation \cite{isobe2020high,PhysRevApplied.13.024053,PhysRevB.102.245422,kumar2021room,he2021quantum,gao2021second,zhang2021terahertz}.

Here, we study the nonlinear Hall effect (NLHE) in two-dimensional electron gases (2DEGs) under an applied inhomogeneous magnetic field such that its total flux through the system is zero.
First, we consider an isotropic 2DEG near the band edge under a magnetic field which is odd with respect to a single boundary. This kind of system has received in the past strong interest due to the particularities of the states formed at the boundary, the so-called snake states \cite{PhysRevLett.68.385,PhysRevLett.72.1518,reijniers2000snake,NatCommsSorbit}.
Based on the quantum kinetic approach, we show that an AC electric field applied perpendicular to the boundary drives a purely nonlinear Hall current.
This is found to be the case also for less symmetric 2DEGs provided they have at least a reflection symmetry.
Second, we study an array of such boundaries where regions of width $2W$ of a sign-changing magnetic field are separated by regions without magnetic field, similar to already constructed devices \cite{weiss1989magnetoresistance,ye1997magnetoresistance,PhysRevB.70.115303,nogaret2010electron}.
We name such configuration a \textit{real-space magnetic field dipole}. By means of semiclassical calculations, we resolve in time and space the Hall current finding that it mainly arises from the snake orbits.
Remarkably, we reveal a parametric resonance controlled by the magnetic profile length scales $L, W$ and the cyclotron radius $R_c$. Consequently, the effect found here is highly tunable, and particularly, by changing the ratio $W/R_c$, the NLHE can be varied by orders of magnitude.

	This work is organized as follows. Section \ref{sec_QKT} analyzes the Hall response via the quantum kinetic theory for the case of a 2DEG with an applied magnetic field odd with respect to a single boundary. Sections \ref{sec_SEMI} and \ref{sec_resonant} study a periodic array of such boundaries via numerical semiclassical calculations with focus on the roles played by different length-scales associated with the magnetic field profile. Section \ref{conclusions} presents our conclusions. 

\section{Quantum kinetic theory for a real-space dipole}
\label{sec_QKT}

\begin{figure}[t]
  \includegraphics[width=0.9\linewidth]{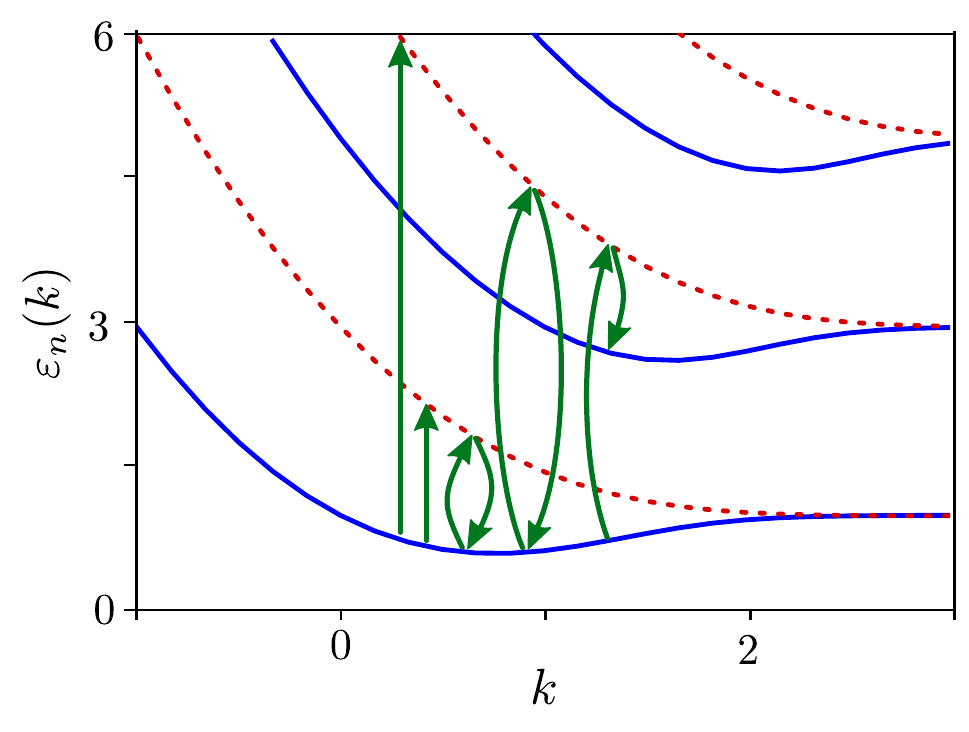}
	\caption{Electronic structure for a 2DEG with vector potential $A(x)=|x|$. Continuous blue (dashed red) lines correspond to Landau levels of even (odd) parity. Vertical straight arrows illustrate optical processes that contribute to the first-order density matrix $\hat{\rho}_{1}$ while pairs of curved arrows those that contribute to the second-order density matrix.
	}
  \label{landau}
\end{figure}

We consider a 2DEG near the band edge and under a perpendicular magnetic field $B(x)$ with an odd profile with respect to a baseline $x=0$.
In the presence of an in-plane AC electric field along $\hat{\bf x}$ of frequency $\omega$, $E_x(t)$, the Hamiltonian reads
\begin{align}
\hat{H} &= \hat{H}_0+e E_x(t) \,\hat{ x} , \\
\hat{H}_0 &= \frac{1}{2m} \hat{p}_x^2 + \frac{1}{2m}\Big[\hat{p}_y-\frac{e}{c}\hat{A}(x)\Big]^2. \label{eq:H0}
\end{align}
The vector potential is chosen in the Landau gauge along $\hat{\bf y}$ as $\hat{A}(x)=\int^x dx'B(x')$ and is an even function of $x$ due to the odd parity of the magnetic field.
We identify the eigenstates of $\hat{H}_0$ as $|k,n\rangle$, with $k$ the momentum along $\hat{\bf y}$ ($[\hat{H}_0,\hat{p}_y]=0$) and $n$ a Landau level index. Matrix elements of an operator $\hat{O}$ are written $\hat{O}_{knn'}=\langle k,n | \hat{O} | k n' \rangle$. The equilibrium wave functions have the form $\psi_{k,n}({\bf r}) = e^{i k y}\phi_{k,n}(x)$
where the functions $\phi_{k,n}(x)$ obey the Schr\"{o}dinger equation
\begin{equation}
\Big[-\frac{d^2}{d^2 x} + V_k(x)\Big] \phi_{k,n}(x) = \varepsilon_{k,n}\phi_{k,n}(x)
\end{equation}
with $V_k(x)$ the effective potential $V_k(x) = \frac{1}{2}\big[ A(x)-k\big]^2$.
Since $V_k(x)$ is even, the functions $\phi_n(x)$ have a defined parity.
For illustration, Fig. \ref{landau} shows the spectrum corresponding to the case $A(x)=|x|$.

To study nonequilibrium properties, we exploit the quantum Liouville equation for the disorder averaged density matrix $\hat{\rho}$ in the relaxation time approximation, which can be written as
\begin{equation}
i\hbar \:\partial_t  \hat{\rho}  -[\hat{H},\hat{\rho}] = -i\hbar\Gamma\big[\hat{\rho}(t)-\hat{\rho}_0\big],
\label{eq_lio}
\end{equation}
where $\hat{\rho}_0$ is the equilibrium density matrix.
While \revision{the relaxation time approximation} has often been used when studying both the linear \cite{PhysRevB.84.205327} and the nonlinear Hall response \cite{inti_prl_2015,Sodemann2019}, 
in order to present the approximations involved we derive Eq. (\ref{eq_lio}) in Appendix \ref{app_rela} relying on the more general framework established in Ref. \cite{Culcer2010,Culcer2017}.
Here, we summarize key aspects \revision{of this derivation}.  
First, collisions can generally mix different terms of the density matrix, yielding a more complex \revision{right hand side of} Eq. (\ref{eq_lio}) which depends on the microscopic details associated with the scattering mechanisms. For weak disorder, the impurity scattering between different Landau levels can be neglected and the collision term is described by the relaxation rate $\Gamma_{kn_1n_2}$, which generally depends on momentun and on Landau level indexes. In this level of approximation, different elements of the density matrix relax towards equilibrium with a different time scale. As explained in Appendix \ref{app_rela}, the additional approximation associated with Eq. $(\ref{eq_lio})$, $\Gamma_{kn_1n_2}\equiv\Gamma$, does not affect the results presented in this Section.
\par
In addition, the quantum kinetic Eq. \eqref{eq_lio} also relies on the Markov approximation for the scattering, which is justified in the limit in which both the frequency of AC electric field $E_x(t)$ and the relaxation rate $\Gamma$ are  smaller than the characteristic frequency $1/\tau_{a}\sim\bar{\varepsilon}/\hbar$, where $\bar{\varepsilon}$ denotes the typical energy of electrons which participate in transport \cite{vasko2006quantum,kohn-luttinger}. 
 Since such energy is in the order 10 to 100\,meV depending on details of the system \cite{koc2003,gra2001}, the following derivation is valid for a weakly disordered 2DEG \cite{ish2020,sch2018,shi2002} with an applied electric field in the THz or lower frequency range ( $\Gamma, \omega \lesssim 10^{12}\, {\rm s}^{-1}$).


Following the well-established perturbation theory
the density matrix is expanded as $\hat{\rho} = \sum_l\hat{\rho}_l$,
with $\hat{\rho}_{l}$ of order $l$ in the electric field \cite{kohn-luttinger,Xiao_2018,PhysRevB.100.165422,Sodemann2019}.
Separation of Eq. (\ref{eq_lio}) in terms of different order yields the recursive set of equations
\begin{align}
	& (0^{\rm th}): \quad 	i\hbar\: \partial_t \hat{\rho}_0  - [ \hat{H}_0 ,   \hat{\rho}_0    ] = 0,  \label{rho-0} \\
	& (l^{\rm th}): \quad 	i\hbar \:\partial_t \hat{\rho}_l   - [ \hat{H}_0 ,   \hat{\rho}_{l}    ] - e E_x [ \hat{x} , \hat{\rho}_{l-1}   ]
	= -i \hbar\Gamma \hat{\rho}_{l},  \label{rho-n}
\end{align}
It follows from Eq. \eqref{rho-0} that $\hat{\rho}_0$ is time-independent and then commutes with $\hat{H}_0$ yielding  $\hat{\rho}_{0,knn'}= \delta_{n,n'} f(\eig)$ where $f$ is the Fermi distribution function.
Next, we decompose $\hat{\rho}_l$ in different harmonics of the applied driving field.
From the time dependence of Eq. (\ref{rho-n}), it can be observed that $\hat{\rho}_1$ contains terms oscillating at $\pm\omega$, $\rho_{1}^\omega$, while  $\hat{\rho}_{2}$ is composed of a constant in time part, $\rho_{2}^0$, plus terms oscillating at $\pm2\omega$, $\rho_{2}^{2\omega}$.
The solution reads
\begin{align}
	\rho_{1,kn_1n_2}^\omega &= -e  \frac{E_x[\hat{x},\hat{\rho}_0]_{kn_1n_2}}{\hbar(\omega-i\Gamma)+\varepsilon_{k,n_1}-\varepsilon_{k,n_2}}
\nonumber 
	\\
	\rho_{2,kn_1n_2}^0 &= -\frac{e}{4} \frac{E_x^*[\hat{x},\hat{\rho}_{1}^{\omega}]_{kn_1n_2} + E_x[\hat{x},\hat{\rho}_1^{\omega\dag}]_{kn_1n_2}}{\varepsilon_{k,n_1}-\varepsilon_{k,n_2}-i\hbar\Gamma} \label{sol_2}\\
	\rho_{2,kn_1n_2}^{2\omega} &= -\frac{e}{2} \frac{E_x [\hat{x},\hat{\rho}_1^\omega]_{kn_1n_2}}{\hbar(2\omega-i\Gamma) +\varepsilon_{k,n_1}-\varepsilon_{k,n_2}}
\nonumber 	
\end{align}
These equations exhibit the same structure than the perturbed distribution functions obtained in the Boltzmann formalism in Ref. \cite{inti_prl_2015} (Eq. (6) therein), with the role played there by derivatives of $f$ along the electric field direction replaced here by commutators $e E_x [\hat{x},\hat{\rho}_{l-1}]$. These measure the amplitude of the transitions induced by the electric field for a given distribution of states $\hat{\rho}_{l-1}$. The linear perturbation $\hat{\rho}_1$ has matrix elements $[\hat{x},\hat{\rho}_0]_{kn_1n_2} =  \hat{x}_{kn_1n_2} [f(\varepsilon_{kn_1})-f(\varepsilon_{kn_2})] $, from where it can be seen that it only mixes Landau levels of opposite parity, otherwise $\hat{x}_{kn_1n_2}$ vanishes. The opposite holds for $\hat{\rho_2}$, which arises from pairs of electric-field induced transitions and, therefore, only mixes states of same parity (see Fig. \ref{landau}).

The Hall current can be obtained summing order by order as
\begin{equation}
	\langle j_{y} \rangle = \sum_l\text{Tr}( {\hat v}_{y}\, \hat{\rho}_{l}) = \sigma_{yx}E_x + \chi_{yxx} E_x^2 + \cdots.
	\label{eq:hall_current}
\end{equation}
The Hall conductivity tensors $\sigma_{yx}$, $\chi_{yxx}$, and so on can be decomposed into different harmonics stemmed from ${\hat{\rho}_l}$'s.
The $+\omega$-component of the linear Hall conductivity  results $\sigma^\omega_{yx}=-e^2\sum_{k,n_1,n_2}
\hat{v}_{y;kn_2n_1} \,
	\hat{x}_{kn_1n_2} \,
	{\cal W}^{\omega}_{kn_1n_2}
$ where ${\cal W}^{\omega}_{kn_1n_2}=(f_{k,n_1}-f_{k,n_2})/(\hbar\omega-i\hbar\Gamma+\varepsilon_{k,n_1}-\varepsilon_{k,n_2})$ and for the opposite harmonic we have $\sigma^{-\omega}_{yx}=(\sigma^\omega_{yx})^\ast$. We thus see that that $\sigma_{yx}$ is zero for the setup proposed here: $\hat{x}_{kn_1n_2}$ vanishes when $n_1$ and $n_2$ have   the same parity and so does $\hat{v}_{y;kn_2n_1}$ when they do not. The latter is due to $\hat{v}_y$ being even with respect to $x$, a condition that remains true for a non-isotropic 2DEGs as long as it presents a reflection symmetry with respect to a line within the 2D plane. In this case, the magnetic field profile should be parallel to this line.

For the second-order Hall conductivity we obtain
\begin{align}
&
	\chi^{2\omega}_{yxx} = -\frac{e^3}{4}\sum_{k,n_i}
	\hat{v}_{y;kn_2n_1} \,
	\hat{x}_{kn_1n_3} \,
	\hat{x}_{kn_3n_2} \,
	{\cal W}^{2\omega}_{k n_1 n_2 n_3},\label{chi2omega}\\
&
	{\cal W}^{2\omega}_{k n_1 n_2 n_3}=
	\frac{{\cal W}_{kn_1n_3}^\omega - {\cal W}_{kn_3n_2}^\omega }{\hbar(2\omega-i\Gamma)+\varepsilon_{k,n_1}-\varepsilon_{k,n_2}},
\end{align}
and $\chi^{-2\omega}_{yxx}=(\chi^{2\omega}_{yxx})^\ast$.
The zeroth harmonic $\chi^{0}_{yxx}$ which describes rectification effects can be obtained replacing ${\cal W}^{2\omega}_{k n_1 n_2 n_3}$ in Eq. \eqref{chi2omega}  by
\begin{align}
	{\cal W}^{0}_{k n_1 n_2 n_3}=
	\frac{{\cal W}_{kn_1n_3}^\omega - {\cal W}_{kn_3n_2}^\omega }{-i\hbar\Gamma+\varepsilon_{k,n_1}-\varepsilon_{k,n_2}}+\{\omega\to-\omega\}.
\end{align}
The terms with $n_1$ and $n_2$ of same parity and opposite to that of $n_3$ make the second-order Hall effect nonzero.
Assuming the limit in which only the lowest Landau is thermally occupied together with a resonance condition $\omega\sim\varepsilon_{n,k}-\varepsilon_{0,k}$ for some range of $k$ (with $n$ some odd-parity Landau level), we find
$\chi_{yxx}^{0}\sim i(e^3/8\hbar^2\Gamma^2)\,
\hat{v}_{y;k,0,0} \,
	|\hat{x}_{k,0,n}|^2$ and $\chi_{yxx}^{2\omega}\sim (\Gamma/\omega)\chi_{yxx}^{0}$.
Note that in related problems the divergence in the limit $\Gamma \to 0$ has been found to become finite by treating the problem beyond the perturbation theory \cite{matsyshyn2021rabi}.

A proper Hall current must be dissipationless \cite{RevModPhys.82.1539}.
In a two-dimensional electron system, the power dissipation attributed to the second-order transverse conductivity reads
\begin{align}
{\bf E}\cdot{\bf j}_{\rm NLH}= \sum_{\alpha\neq\beta}\big( \chi^{0}_{\alpha \beta\beta} + \chi^{0}_{\beta\alpha\beta}\big|^{\ast} +\chi^{0}_{\beta\beta\alpha } \big|^{\ast} \big) E_\alpha^\ast E_{\beta}^2
\label{eq:dissipation}
\end{align}
We have verified explicitly in Appendix \ref{app_diss} that there is no power dissipation for the second-order Hall current obtained from Eq. (\ref{eq:hall_current}), in the limit of $\Gamma \to 0$.

\begin{figure*}[t]
  \includegraphics[width=0.98\textwidth]{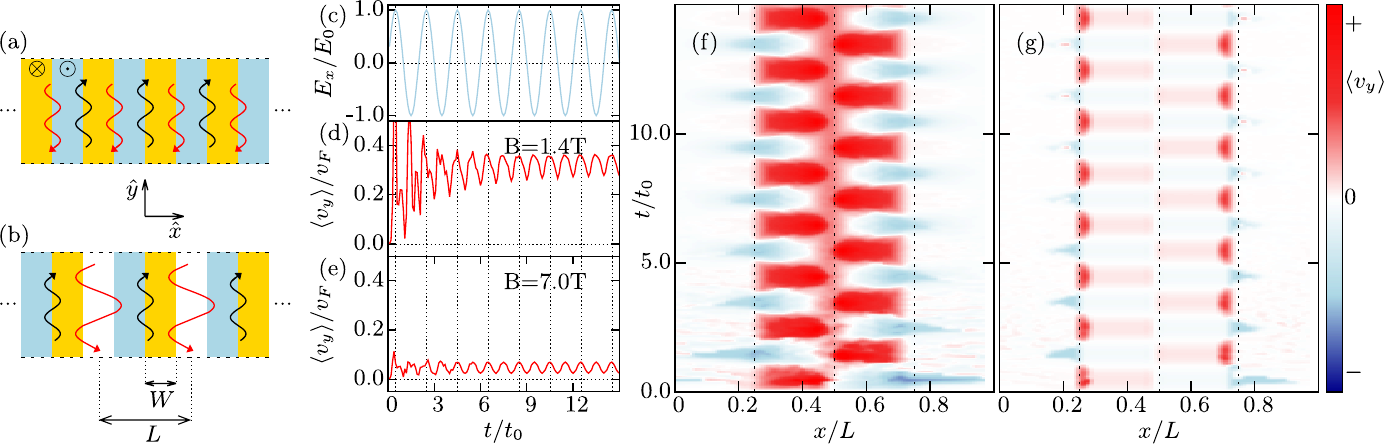}
	 \caption{(a) $S$-symmetric magnetic field profile $B_z (x)$. Blue and yellow fields correspond to negative ($\odot$) and positive ($\otimes$) magnetic field, respectively.
  (b) $S$-broken configuration.
  (c) Applied electric field along $\hat{x}$.
  (d,e) Hall current for different magnetic field amplitudes, $B=1.4\,$T and $7.0\,$T.
	 (f,g) Time- and space-resolved Hall current for $B=1.4\,$T and $7.0\,$T. Vertical dashed lines indicate the positions at which the magnetic field changes.
 $E_0=10^{-2}\,$V$/$nm, $L=10^3\,$nm and $t_0=L/v_F=1/$THz.}
  \label{fig:2}
\end{figure*}

The nonlinear anomalous Hall effect arises from the anomalous velocity which is itself proportional to the electric field \cite{inti_prl_2015}. Due to this, the second-order response is determined by the first-order perturbation of the distribution function. In this sense, the electric field plays two roles, giving rise to the anomalous velocity of carriers and perturbing the distribution function such that the average anomalous velocity is finite. Importantly, the latter also requires the point symmetry to be sufficiently low, in particular, inversion symmetry must be broken.
The effect here described is different. The crucial symmetry is the even parity of $H_0$ with respect to $x$, which in turn is inherited by $\hat{v}_y$.
This forces the linear Hall response to vanish: To this order, only states of opposite parity are connected by the electric field-induced transitions but $\hat{v}_{y}$ has zero overlaps between such states hence these transitions do not yield a Hall current.
The opposite is true in the second-order response: $\hat{\rho}_2$ mixes Landau levels of the same parity which can hold a Hall current.

\section{Semiclassical theory}
\label{sec_SEMI}


We now consider the semiclassical limit. We assume the 2DEGs to be inversion- and $\Theta$-symmetric in the absence of external fields.
The equations of motion for weak external fields are
  \begin{align}
    \dot { \vvr }       &= \vvv_\vvk  = \frac{1}{\hbar} \PPvvk{\energy(\vvk)} \label{eq_sm1} \\
   {\hbar} \dot{ \vvk }  &= e\, \big[\,  \vvE ( t ) + \frac{1}{c} \vvv_\vvk \times \vvB ( x ) \,\big]
\label{eq_sm}
  \end{align}
where $\vvr$ and $\vvk$ represent wave-packet center of mass in real and momentum spaces, respectively \cite{PhysRevB.53.7010}.
We omit the band indices since we consider a homogeneous 2DEG near the band edge and we assume a quadratic energy dispersion. Usually, these equations of motions are supplemented with a Boltzmann approach to obtain the nonequilibrium distribution function \cite{mahan2013many,PhysRevB.59.14915,PhysRevLett.112.166601}. Here, we follow a distinct approach, useful in the clean limit $\Gamma \to 0$, which consists of computing the Hall current based on the numerical integration of the trajectories followed by independent particles governed by Eqs. (\ref{eq_sm1}-\ref{eq_sm}).
This approach amounts to numerically sampling the distribution function in the clean limit and allows us to compute the Hall current including all of its harmonics and not necessarily close to the weakly nonlinear regime \cite{Rostoker1964}.

Before presenting the numerical results, it is instructive to analyze within this classical limit why the linear Hall effect vanishes while higher-order effects are expected to be nonzero.
Using the Einstein relation between the diffusion and the conductivity tensors, the former can be written in terms of correlations in time of different components of the velocity,
\begin{align}
\sigma_{ij} = e^2 N \int_{0}^{\infty} dt \: \big\langle v_i(t) v_j(0) \big\rangle
\end{align}
where $i,j=\{x,y\}$ and $N$ is the density of states at the Fermi energy \cite{Beenakker1991}.
The brackets $\langle...\rangle$ denote an average over the available phase space.
The linear response is determined by the correlator $\langle v_i(t) v_j(0) \rangle$ in the absence of electric field.
For quadratic bands we can formally integrate Eq. \eqref{eq_sm} and obtain the velocity to zeroth-order in $E_x$ as
\begin{align}\label{eq:vy0}
	v^{(0)}_x (x,t) &= v_{x,0} + \frac{e}{mc} \int_{\mathcal{P}} dy(t)\,B[x(t)]
, \\
\label{eq:vx0}
	v^{(0)}_y (x,t) &= v_{y,0} - \frac{e}{mc} \big\{A[x(t)] - A [x(0)] \big\}  ,
\end{align}
Here, $v_{x,0}$ and $v_{y,0}$ are the initial (random) velocities, $\mathcal{P}$ the path associated with the particle trajectory and the identity $\int dt \: {\bf v} \, B(x) \equiv \int d{\bf r} \: B(x)$ has been used. 
Therefore, due to parity properties of $B(x)$ and $A(x)$,  $v^{(0)}_x (x,0)$ and $v^{(0)}_y (x,t)$ are odd and even functions of $x$, respectively, leading to a vanishing first-order Hall conductivity, this time from a semiclassical perspective.

Last, a consideration of the correlator $\langle v_i(t) v_j(0) \rangle$ including a correction of the velocities due to the applied electric field already hints towards the existence of higher-order Hall effect.
The velocity corrections read
\begin{align}
v^{(1)}_x(x,t) &= \frac{e}{m} \int^t dt\: E_x \\
v^{(1)}_y(x,t) &= -\frac{e}{mc} \int^t dt \:\vx{(1)}(x,t) B(x) \label{eq:dotVy1}
\end{align}
and yield a contribution to the second-order Hall conductivity proportional to $\langle v^{(1)}_y(x,t)\, v_x^{(0)}(x,0) \rangle$. Since both $v^{(1)}_y( x ,t )$ and $v^{(0)}_x( x, 0)$ are odd functions of $x$, this contribution is nonzero.

\section{Parametric resonance and higher harmonics generation}
\label{sec_resonant}

We now present numerical results for a time-dependent electric field $\vvE ( t )=E_0 \sin(\omega t) \hat{x}$,
with $\omega$ in the THz range, 
obtained by numerically integrating Eqs. (\ref{eq_sm1}-\ref{eq_sm}) with the Runge–Kutta method. 
We consider $2\times 10^6$ particles being initially uniformly distributed in space and having as initial conditions the Fermi velocity $v_F=10^{6}\,$m/s and the Fermi wave vector $10^{-2}\, 1/{\rm\AA}$  \cite{Beenakker1991}.
The Hall current is associated with the average velocity $\langle v_y\rangle$.

We consider periodic boundary conditions so that the system can be regarded as an array of magnetic field steps, as shown in Fig. \ref{fig:2}(a),(b). In order to obtain a Hall response, the system must not be symmetric under $S=\Theta\times T_{1/2}$, with $T_{1/2}$ a translational vector along $\hat{x}$.
This symmetry forces $j_y$ to vanish to all orders in the electric field $E_x$. We show this using the quantum kinetic approach and we also recover this result numerically in our semiclassical calculations (see Appendix \ref{app_symm}).
One possibility to break $S$ is the inclusion of regions where no magnetic field is applied, as shown in Fig. \ref{fig:2}(a),(b). For an $S$-symmetric system, snake orbits of neighboring boundaries exactly mirror each other. When $S$ is broken, a finite real-space magnetic-field dipole arises, which reflects in the asymmetry of neighboring snake orbits. Note that, following the previous sections, a single unit cell in the array is expected to have a finite purely nonlinear Hall effect and in a finite array the net Hall conductance is proportional to the number of unit cells.

Fig. \ref{fig:2}(c) shows the applied electric field while panels (d) and (e) show the resulting Hall current $\langle v_y \rangle$ for two different magnetic fields. After a transient time, the Hall current becomes approximately periodic and is characterized by a DC offset and an AC THz component twice faster than the driving electric field. A Fourier analysis of the signals shows that they are governed by only even multiples of the driving frequency (see Appendix \ref{app_fourier}). 
This result obtained in the semiclassical limit is consistent with that in the quantum limit and together 
firmly establish the possibility of NLHE and second harmonic generation in inversion-symmetric 2DEGs under an inhomogeneous magnetic field.

It is instructive to resolve in time and space the Hall current.
Results for the chosen magnetic fields in panels (d) and (e) are shown in (f) and (g), respectively.
These show, first, that regions without applied magnetic field while being essential to break the $S$ symmetry do not contribute significantly to $\langle v_y\rangle$, which is expected since electrons in these regions are only driven by the electric field. 
Second, while which area contributes the most can indeed depend noticeably on the magnetic field strength, high contributions tend to locate near the lines at which magnetic field changes sign, hinting to a phenomenon truly arising from the boundary and, therefore, different from previous mechanisms for NLHE that originate in the bulk electronic structure.  
In particular, the oscillating pattern in panel (f) reflects the formation of snake orbits and indicates that these provide the main contribution to the Hall current.

\begin{figure}[t]
	\includegraphics[width=\linewidth]{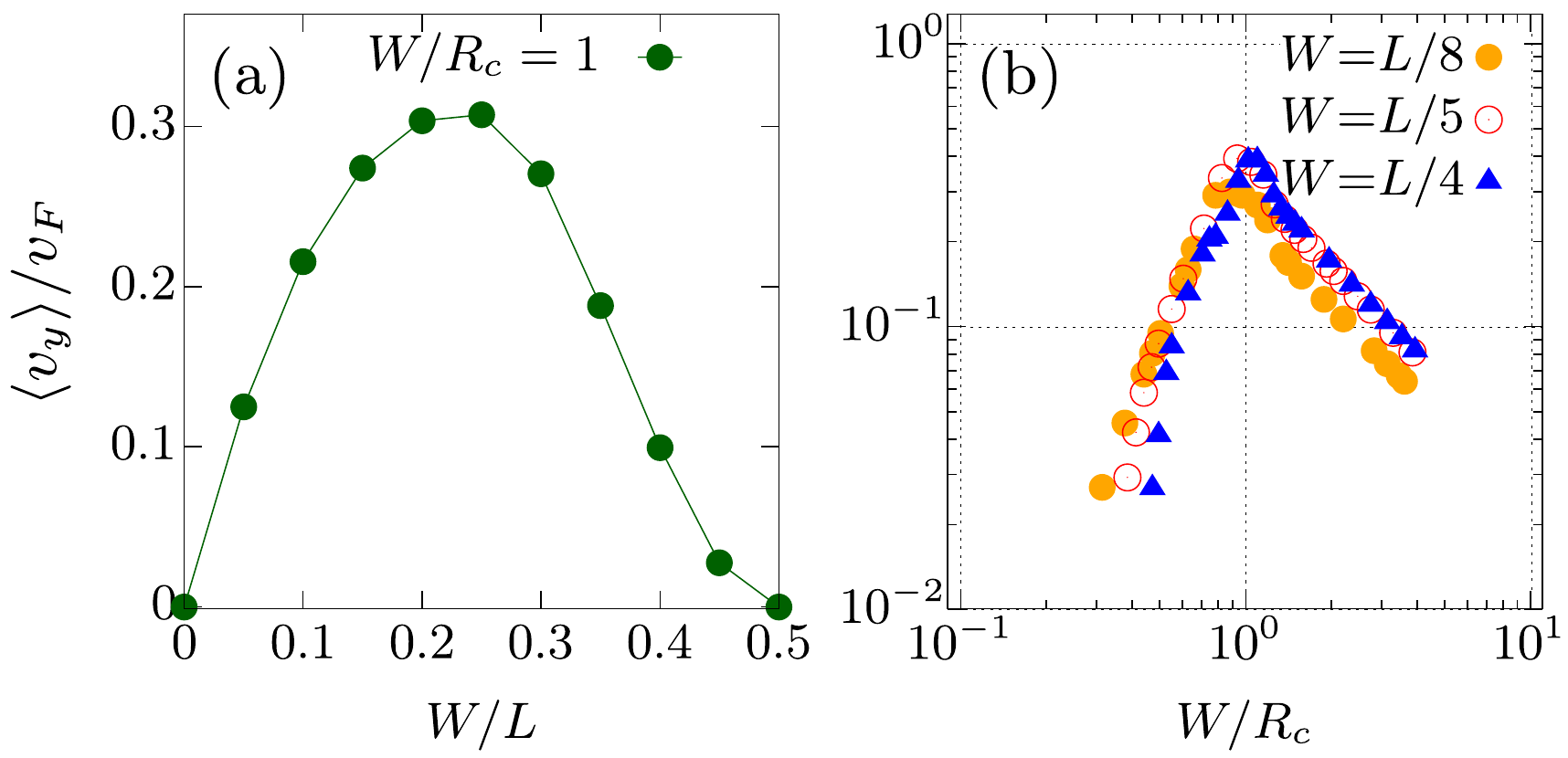}
	\caption{Hall current as a function of $W/L$ (a) and of $W/R_c$ (b).  $L=10^{3}\,$nm and $E_0=10^{-3}$\,V/nm.}
  \label{fig3:}
\end{figure}

We now analyze the role played by different length scales associated with the magnetic field profile, 
 assuming the mean free path to be the largest length. 
Fig. \ref{fig3:}(a,b) show the Hall current as a function of $W/L$ and of $W/R_c$, respectively.
Both curves display a maximum, which together reveal how the NLHE can be engineered via characteristic lenght scales.
Naturally, a finite $W$ is required so the Hall current vanishes when $W\to0$.
Increasing $W/L$ is beneficial, as long as $W/L<1/4$. Beyond this point, the system approaches the limit $W/L\to1/2$   where it vanishes due to the restoration of the $S$-symmetry.
For a wide range of parameters around $W/L\sim1/4$, 
the resulting Hall current is largely controlled by $W/R_c$, as indicated by the nearly collapse of the data obtained from different values of $W/L$ shown in Fig. \ref{fig3:}(b).
The strong susceptibility to the ratio $W/R_c$ reveals a parametric resonance where the NLHE is maximized when the effective width of the snake orbits approaches the maximum value they can have.
This optimum condition corresponds to the profile shown in Fig. \ref{fig:2}(f).
 
The resonant NLHE can be experimentally explored in a large variety of systems. Devices with a corrugated magnetic field that changes its sign have been experimentally established in AlGaAs-GaAs heterojunctions in the proximity of Dy \cite{ye1997magnetoresistance} or Co \cite{nogaret2010electron} stripes.
The parameters chosen for the simulations are all in the order of magnitude of well-established 2DEGs such as AlGaAs-GaAs and Si \cite{Beenakker1991}.
Last, the condition of having a reflection symmetry to which the magnetic profile should be aligned in order to ensure a vanishing linear Hall conductivity is satisfied by many two-dimensional systems including graphene \cite{novoselov2004electric}, transition metal dichalcogenides \cite{PhysRevLett.105.136805,PhysRevLett.108.196802,PhysRevB.88.085440} and oxide interfaces \cite{ohtomo2004high,brinkman2007magnetic,PhysRevB.86.125121,PhysRevLett.110.206805}.

\
\section{Conclusions}
\label{conclusions}

In summary, we have established theoretically that a purely nonlinear Hall response can be produced in two-dimensional electron gases by engineering the magnetic field profile. 
Different from previously explored mechanisms that originate in the bulk electronic structure, the effect here described stems from states formed at a boundary in which the magnetic field changes sign. Lattice inversion symmetry breaking is not required, only the existence of a mirror symmetry to which the magnetic profile should be aligned is important for the vanishing of the linear response. 
Last, accesible length scales set by the magnetic field profile provide a clean way of engineering the nonlinear Hall response.

\begin{acknowledgments}
We are thankful to Cosma Fulga, Jhih-Shih You, Sheng-Chin Ho, Tse-Ming Chen, Carmine Ortix and Jeroen van den Brink for enlightening discussions.
A.G.M. acknowledges financial support from Iran Science Elites Federation under Grant No. 11/66332.
J.I.F. would like to thank the support from the Alexander von Humboldt Foundation. C.-H.C. acknowledges the financial support by the Ministry of Science and Technology (grant numbers MOST-107-2112-M-006-025-MY3 and MOST-108-2638-M-006-002-MY2) and by the Yushan Young Scholar Program under the Ministry of Education (MOE) in Taiwan. 
\end{acknowledgments}

\appendix

\section{Relaxation time approximation}
\label{app_rela}

We start from the quantum Liouville equation \cite{Culcer2010, Culcer2017}
\begin{equation}
i\hbar \:\partial_t  \hat{\rho}  -[\hat{H},\hat{\rho}] = -i\hbar\: {\cal I}(\hat{\rho}).
\label{eq_dis_ave}
\end{equation}
Here, the collision operator ${\cal I}(\hat{\rho})$ accounts for the average effect on the density matrix of the scattering with an impurity potential $\hat{U}$.  Within the Born approximation, it reads
\cite{Sodemann2019}
\begin{widetext}
\begin{align}
{\cal I}(\hat{\rho})|_{kn_1n_4}=\frac{\pi {\cal N}_{\rm imp}}{\hbar} \sum_{k'n_2n_3} \Big[
&
{\cal U}^{n_1n_2}_{kk'} \: {\cal U}^{n_2n_3}_{k'k}  \rho_{kn_3n_4} \:\delta(\varepsilon_{k'n_2}-\varepsilon_{kn_3})
+
\rho_{kn_1n_2} \: {\cal U}^{n_2n_3}_{kk'} \: {\cal U}^{n_3n_4}_{k'k}   \:\delta(\varepsilon_{k n_2}-\varepsilon_{k'n_3}) \nonumber\\
-&
{\cal U}^{n_1n_2}_{kk'}  \rho_{k'n_2n_3}
\: {\cal U}^{n_3n_4}_{k'k}  \:\delta(\varepsilon_{k'n_3}-\varepsilon_{kn_4})
-
{\cal U}^{n_1n_2}_{kk'}  \rho_{k'n_2n_3}
\: {\cal U}^{n_3n_4}_{k'k} 
\:\delta(\varepsilon_{k n_1}-\varepsilon_{k'n_2})
\Big],
\label{eq_col}
\end{align}
\end{widetext}
where ${\cal U}^{n_1n_2}_{kk'} = \langle kn_1| U({\bf r}) |k'n_2\rangle $ and ${\cal N}_{\rm imp}$ is the impurity density. 
	In the weak disorder limit and assuming a smooth impurity potential profile, the impurity scattering between Landau levels can be neglected  and only the forward scattering terms ${\cal U}^{nn}_{kk}\equiv {\cal U}_{kn}$ are considered.
Then, introducing the velocity  $\vartheta_{kn}=\partial_{k}\varepsilon_{kn}$, the collision operator simplifies to 
\begin{align}
	{\cal I}(\hat{\rho})|_{kn_1n_2}&=\frac{\pi {\cal N}_{\rm imp}}{\hbar} 
    \Big[
    \frac{{\cal U}_{kn_1}^2 }{\vartheta_{kn_1}}
    +
    \frac{{\cal U}_{kn_2}^2
    }{\vartheta_{kn_2}} \nonumber\\
    &- 
    {\cal U}_{kn_1}\,{\cal U}_{kn_2}
    \big(     \frac{1}{\vartheta_{kn_1}}
+     \frac{1}{\vartheta_{kn_2}}
 \big) \Big]\, \rho_{kn_1n_2} \label{eq_collision0}\\
 &=\Gamma_{kn_1n_2}  \rho_{kn_1n_2}, 
 \label{eq_collision}
\end{align}
	where we have defined the relaxation rate $\Gamma_{kn_1n_2}$ as the prefactor of $\rho_{kn_1n_2}$ in the collision term. 
$\Gamma_{kn_1n_2}$ encodes microscopic information that measure the time scale in which the density matrix element $\rho_{kn_1n_2}$ relaxes to equilibrium due to the scattering with impurities.
It naturally vanishes when $n_1=n_2$, ensuring that the collision operator is zero when evaluated at a diagonal density matrix such as the equilibrium density matrix $\hat{\rho}_0$.
Generally, it follows from Eq. (\ref{eq_collision0}) that, given two states  $|k,n_1\rangle$ and $|k,n_2\rangle$, the more different are their band velocities and the more different are their couplings to the impurity potential, the slowest the relaxation of $\rho_{kn_1n_2}$ is. 
 These microscopic details are erased 
 by the two additional approximations that yield Eq. (\ref{eq_lio}), namely, neglecting of the relaxation rate dependence on the Landau level indexes and on the momentum.
\par
 Let us briefly comment on their possible effects on our results.
 Once at the level of approximation of Eq. (\ref{eq_collision0}),
  the $\{n_1,n_2\}$-dependence of the relaxation rate is irrelevant for the discussion of the Hall effect in a 2DEG with a single magnetic field boundary presented in Sec. \ref{sec_QKT}. 
 The fundamental reason for this is that, different to Eq. (\ref{eq_col}), in Eq. (\ref{eq_collision0}) the collision operator does not mix different elements of the density matrix. 
 The dependence on momentum is also irrelevant since the derivation in that section does not use the momentum dependence of the density matrix.
 In conclusion, all results in Sec. \ref{sec_QKT} remain intact replacing $\Gamma$ by $\Gamma_{k n_1n_2}$.
 On the other hand, the momentum dependence of $\Gamma$ is implicitly considered in the Appendix \ref{app_symm} and may affect the results of that section if $\Gamma$ is not even in $k$.

\vspace{2mm}
\section{Dissipationless nonlinear Hall current}
\label{app_diss}

In this Section, we show a reciprocity relation satisfied by the second-order Hall conductivity tensor in the limit $\Gamma \to 0$ that implies the vanishing of Eq. \eqref{eq:dissipation}. We leave the case of finite relaxation rate for future study.

First, we note that the matrix elements of the position and velocity operators are related to each other by 
\begin{align}
    \hat{x}_{\alpha;kn_1n_2} =  \frac{i\hbar\, \hat{v}_{\gamma;kn_1n_2}}{\varepsilon_{kn_2}-\varepsilon_{kn_1} } ,
\end{align}
which follows from the Heisenberg equation of motion
\begin{align}
    \hat{v}_{\gamma;kn_1n_2} 
    &= \frac{d}{dt}\hat{x}_{\alpha;kn_1n_2} = \frac{[\hat{x}_\alpha,\hat{H}_{0}]_{kn_1n_2}}{i\hbar}, \\
    &= \frac{d}{dt}\hat{x}_{\alpha;kn_1n_2} = \frac{\hat{x}_{\alpha;kn_1n_2}(\varepsilon_{kn_2}-\varepsilon_{kn_1})}{i\hbar} .
\end{align}
Consequently, the rectification term of the nonlinear Hall response can be re-written as
\begin{align}
    \chi^{0}_{\alpha\beta\gamma} 
    &\propto \sum_{k,n_i}
    \hat{v}_{\alpha;kn_2n_1}\: \hat{x}_{\beta;kn_1n_3}\: \hat{x}_{\gamma;kn_3n_2}\: {\cal W}^{0}_{kn_1n_2n_3}, \\
    &\propto
    \sum_{k,n_i}
    \frac{-\hat{v}_{\alpha;kn_2n_1}\: \hat{v}_{\beta;kn_1n_3}\: \hat{v}_{\gamma;kn_3n_2}}{(\varepsilon_{kn_1}-\varepsilon_{kn_3})(\varepsilon_{kn_3}-\varepsilon_{kn_2})}\:
    {\cal W}^{0}_{kn_1n_2n_3},
\end{align}
In 2D, the transverse components of the second-order conductivity tensor read
\begin{align}
&    \chi^{0}_{\alpha\beta\beta} 
    \propto
 \sum_{k,n_i}\frac{-\hat{v}_{\alpha;kn_2n_1}\: \hat{v}_{\beta;kn_1n_3}\: \hat{v}_{\beta;kn_3n_2}}{(\varepsilon_{kn_1}-\varepsilon_{kn_3})(\varepsilon_{kn_3}-\varepsilon_{kn_2})}\:
    {\cal W}^{0}_{kn_1n_2n_3} , \label{eq:chi-0-abb} \\
&        \chi^{0}_{\beta\alpha\beta} 
    \propto
     \sum_{k,n_i}\frac{-\hat{v}_{\beta;kn_2n_1}\: \hat{v}_{\alpha;kn_1n_3}\: \hat{v}_{\beta;kn_3n_2}}{(\varepsilon_{kn_1}-\varepsilon_{kn_3})(\varepsilon_{kn_3}-\varepsilon_{kn_2})}\:
    {\cal W}^{0}_{kn_1n_2n_3}, \label{eq:chi-0-bab}\\
&        \chi^{0}_{\beta\beta\alpha} 
    \propto
    \sum_{k,n_i}\frac{-\hat{v}_{\beta;kn_2n_1}\: \hat{v}_{\beta;kn_1n_3}\: \hat{v}_{\alpha;kn_3n_2}}{(\varepsilon_{kn_1}-\varepsilon_{kn_3})(\varepsilon_{kn_3}-\varepsilon_{kn_2})}\label{eq:chi-0-bba}\:
    {\cal W}^{0}_{kn_1n_2n_3},
\end{align}
for $\alpha=x,\beta=y$ and vice versa.
By taking the complex conjugates of Eqs. \eqref{eq:chi-0-bab} and \eqref{eq:chi-0-bba}, we find 
\begin{align}
&      \chi^{0}_{\beta\alpha\beta} \big|^{\ast}
    \propto
     \sum_{k,n_i}\frac{-\hat{v}_{\alpha;kn_2n_1}\, \hat{v}_{\beta;kn_1n_3}\, \hat{v}_{\beta;kn_3n_2}}{(\varepsilon_{kn_1}-\varepsilon_{kn_2})(\varepsilon_{kn_2}-\varepsilon_{kn_3})}
     {\cal W}^{0}_{kn_1n_3n_2} \big|^{\ast}, \\
&        \chi^{0}_{\beta\beta\alpha} \big|^{\ast}
    \propto
     \sum_{k,n_i}\frac{-\hat{v}_{\alpha;kn_2n_1}\, \hat{v}_{\beta;kn_1n_3}\, \hat{v}_{\beta;kn_3n_2}}{(\varepsilon_{kn_3}-\varepsilon_{kn_1})(\varepsilon_{kn_1}-\varepsilon_{kn_2})}
    {\cal W}^{0}_{kn_3n_2n_1} \big|^{\ast},
\end{align}
where we have also interchanged the indices $n_2\leftrightarrow n_3$ and $n_1\leftrightarrow n_3$ inside them, respectively.

We also see that ${\cal W}^{0}_{kn_1n_2n_3} \big|^{\ast}={\cal W}^{0}_{kn_2n_1n_3}$, and subsequently, $\chi^{0}_{\beta\alpha\beta} \big|^{\ast}=\chi^{0}_{\beta\beta\alpha} $.
Putting all above results together, we find
\begin{align}
 & \chi^{0}_{\alpha \beta\beta} + \chi^{0}_{\beta\alpha\beta} +\chi^{0}_{\beta\beta\alpha } \propto \sum_{k,n_i}\hat{v}_{\alpha;kn_2n_1}\, \hat{v}_{\beta;kn_1n_3}\, \hat{v}_{\beta;kn_3n_2}\nonumber\\
& \qquad \qquad \qquad \times\big[ 
  \frac{{\cal W}^{0}_{kn_1n_2n_3} }{(\varepsilon_{kn_1}-\varepsilon_{kn_3})(\varepsilon_{kn_3}-\varepsilon_{kn_2})}
  \nonumber\\
  & \qquad \qquad\qquad  +
   \frac{{\cal W}^{0}_{kn_3n_1n_2} }{(\varepsilon_{kn_1}-\varepsilon_{kn_2})(\varepsilon_{kn_2}-\varepsilon_{kn_3})}
   \nonumber\\
   & \qquad \qquad \qquad
   +
    \frac{{\cal W}^{0}_{kn_2n_3n_1} }{(\varepsilon_{kn_1}-\varepsilon_{kn_2})(\varepsilon_{kn_3}-\varepsilon_{kn_1})}
  ],\end{align}
Then, using the Eq. (11), and taking the limit of $\Gamma\to0$, we obtain
\begin{align}
 &   \chi^{0}_{\alpha \beta\beta} + \chi^{0}_{\beta\alpha\beta} +\chi^{0}_{\beta\beta\alpha } 
 \nonumber\\
  & \propto - \sum_{k,n_i} \frac{\hat{v}_{\alpha;kn_2n_1}\, \hat{v}_{\beta;kn_1n_3}\, \hat{v}_{\beta;kn_3n_2}}{(\varepsilon_{kn_1}-\varepsilon_{kn_2})(\varepsilon_{kn_2}-\varepsilon_{kn_3})(\varepsilon_{kn_3}-\varepsilon_{kn_1})}   \nonumber\\
 &  \qquad \times \big[{\cal W}^\omega_{13}-{\cal W}^\omega_{32}+{\cal W}^\omega_{32}-{\cal W}^\omega_{21}+{\cal W}^\omega_{21}-{\cal W}^\omega_{13} \nonumber \\
& \qquad\qquad + \{\omega\to-\omega\}\big]=0,\label{eq:reciprocity-chi}
\end{align}
where the the compact notation ${\cal W}^\omega_{ij}={\cal W}^\omega_{kn_in_j}$ has been used. We notice that the sum of these second-order Hall responses in the limit of $\Gamma\to 0$ vanishes irrespective of the details of the model encoded in the velocity matrix elements $\hat{v}_{\alpha;knn'}$, and that due to the broken-time reversal symmetry each of these terms do not necessarily vanish in such limit, as shown by the example presented in Section \ref{sec_QKT}.
Last, since we have seen that
$\chi^{0}_{\beta\alpha\beta} \big|^{\ast}=\chi^{0}_{\beta\beta\alpha} $, the reciprocity relation \eqref{eq:reciprocity-chi}
can be equivalently written as $\chi^{0}_{\alpha \beta\beta} + \chi^{0}_{\beta\alpha\beta}\big|^{\ast} +\chi^{0}_{\beta\beta\alpha } \big|^{\ast}=0$ and, therefore,  the power dissipation attributed to the second-order transverse conductivity (Eq. (\ref{eq:dissipation})) vanishes.

\section{Hall response under combined time-reversal and translation symmetry}
\label{app_symm}
In this Section, we elucidate that in the presence of an additional symmetry
which is the combination of time-reversal symmetry with a half-period spatial translation denoted by $\hat{ S}= \hat{\Theta} \times \hat{ T}_{1/2}$, the nonlinear Hall responses become identically zero. An example of a magnetic profile with this symmetry is shown in Fig. 2(a). It should be noted that since time-reversal operator $\hat{\Theta}$ changes the direction of the sign of magnetic field and also the momenta, equivalently, we can assume this symmetry as a screw displacement which is a combination of $\pi$-rotation around $x$-axis and half-period spatial translation ($\hat{S}_{\rm screw}= \hat{R}_{\pi,{\bf x}} \times \hat{ T}_{1/2}$).

In the presence of $\hat{ S}$ symmetry, the magnetic field profile should satisfy the relation ${\bf B}(x)=-{\bf B}(x+W)$ which also leads to $A(x)=-A(x+W)+{\rm cte}$. Since any constant term in the vector potential can be dropped as it has no physical effect due to the gauge invariance, we can always consider $A(x)=-A(x+W)$. So, we can deduce the symmetry 
\begin{align}
\hat{ S}\,V_{k}(x)\,\hat{ S}^{-1}=V_{-k}(x+W)=V_{k}(x),
\end{align}
for the effective potential in Eq. (3). Subsequently, we find that
the eigenenergies are symmetric in $k$ as $\varepsilon_{k,n}=\varepsilon_{-k,n}$, and 
for a given eigenstate $\phi_{k,n}(x)$ with energy $\varepsilon_{k,n}$, the corresponding state $\hat{ S}\phi_{k,n}(x)=\phi_{-k,n}(x+W)$ is also an eigenstate.
 
Now based on above symmetry relations, we see that the matrix elements 
of the $\hat{x}$ operator are also symmetric as
\begin{align}
\hat{x}_{k n_1 n_2}&=\int dx \: \phi^\ast_{k,n_1}(x)\: x \:\phi_{k,n_2}(x)\nonumber\\
&=\int dx \: \phi^\ast_{k,n_1}(x)\: 
\hat{ S}^{-1}\,\big(\hat{ S} \,
x \,\hat{ S}^{-1}\big) \,\hat{ S} \:\phi_{k,n_2}(x)\nonumber\\
&=\int dx \: \phi^\ast_{-k,n_1}(x+W)\: (x+W) \:\phi_{-k,n_2}(x+W)\nonumber\\
&=\int dx \: \phi^\ast_{-k,n_1}(x)\: x \:\phi_{-k,n_2}(x)\nonumber\\
&=\hat{x}_{-k n_1 n_2}. \label{eq:x-under-S}
\end{align}
In the same way, we find that the vector potential also satifies
\begin{align}
&A_{k n_1 n_2}=\int dx \: \phi^\ast_{k,n_1}(x)\: A(x) \:\phi_{k,n_2}(x)\nonumber\\
&\quad=
\int dx \: \phi^\ast_{k,n_1}(x)\: 
\hat{ S}^{-1}\,\big[\hat{ S} \,
A(x) \,\hat{ S}^{-1}\big] \,\hat{ S}
 \:\phi_{k,n_2}(x)\nonumber\\
&\quad=-\int dx \: \phi^\ast_{-k,n_1}(x+W)\: A(x+W) \:\phi_{-k,n_2}(x+W)\nonumber\\
&\quad=-A_{-k n_1 n_2}. \label{eq:A-under-S}
\end{align}

From Eq. \eqref{eq:A-under-S} we can readily see that the transverse velocity 
$v_y$ whose matrix elements are given by the gauge-invariant form
\begin{equation}\label{eq:jy_sym}
  \hat{v}_y|_{kn_1n_2}=   \frac{\hbar k }{m} \,\delta_{n_1,n_2}    - \frac{e}{mc} A_{k n_1 n_2},
\end{equation}
is odd under the change of the momentum sign ($k\to-k$).
However, Eq. \eqref{eq:x-under-S} indicates that the position operator matrix elements $\hat{x}_{k n_1 n_2}$ are even with respect to $k$ 
as well as the transition-rate functions such as ${\cal W}_{kn_1n_2n_3}^{2\omega}$ and ${\cal W}_{kn_1n_2n_3}^{0}$.
The latter can be readily seen from the fact that the transition-rate functions only depend on the energies (and not the eigenstates) which remain unchanged under $k\to-k$. Note that this statement does not hold if the relaxation rate $\Gamma$ is not even in $k$.
Therefore, the overall summands in the linear and nonlinear Hall responses (including all orders) are odd with respect to $k$. Therefore, in the presence of $\hat{S}$ symmetry, the contributions of opposite $k$'s in any order cancel out each other and the Hall response identically vanishes.

\section{Fourier Analysis.}
\label{app_fourier}

\begin{figure}
	\includegraphics[width=0.9\linewidth]{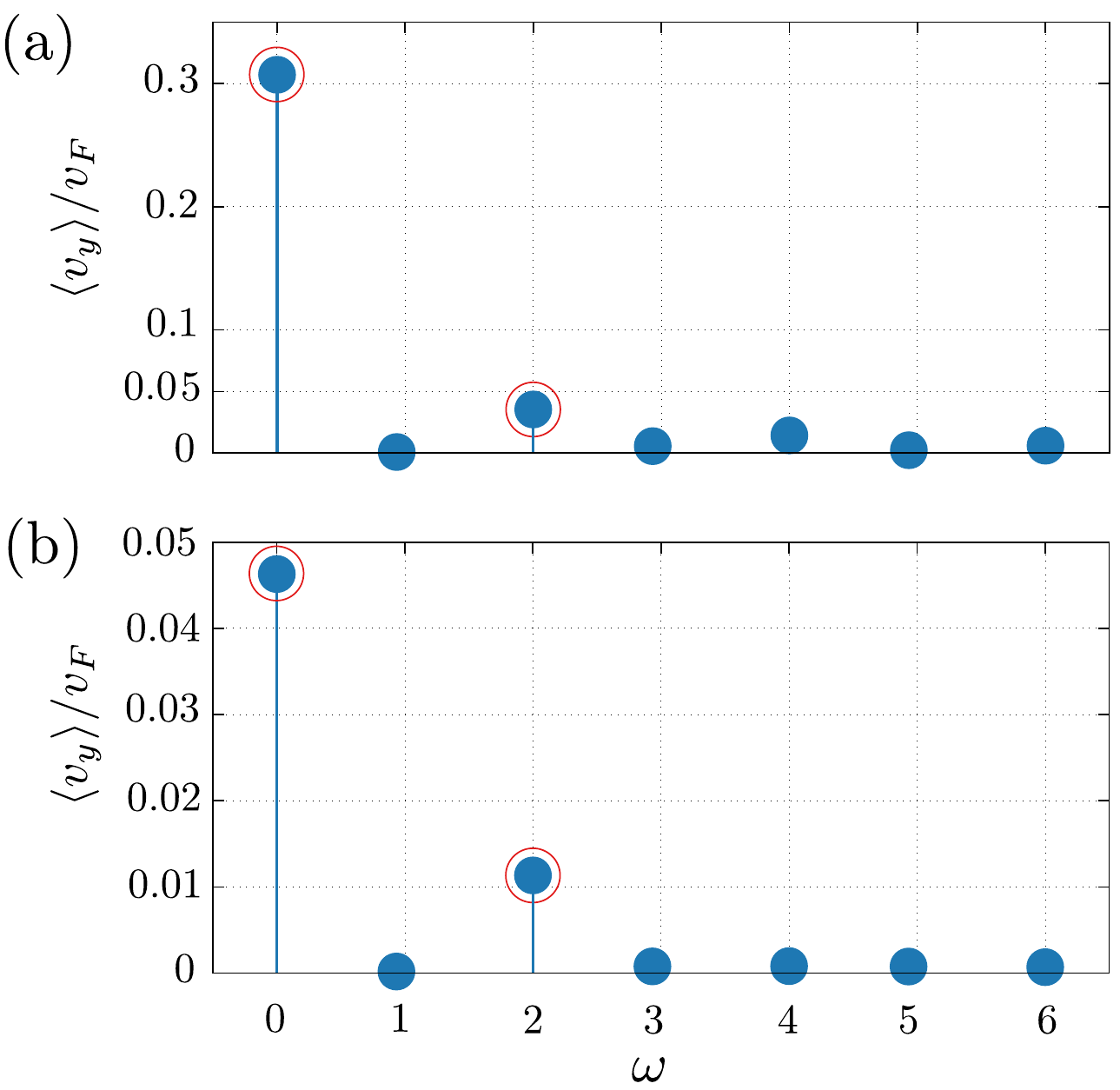}
	\caption{Fourier transform of the average velocity. 
	In panels (a) and (b) the parameters are chosen as those for Fig. 2(d) and 2(e), respectively.}
  \label{figure}
\end{figure}

Figure \ref{figure} shows the Fourier transform of the Hall current presented in Figs. 2(d) and 2(e), respectively. As expected, only the even harmonics of the driving frequency are finite.

\bibliography{ms}

\end{document}